\documentstyle[12pt]{article}
\topmargin = - 1 cm
\font\twelve=cmbx10 at 15pt
\font\ten=cmbx10 at 12pt

\renewcommand{\thefootnote}{\fnsymbol{footnote}}

\begin{document}

\begin{titlepage}

\begin{center}

{\ten Centre de Physique Th\'eorique\footnote{Unit\'e Propre de
Recherche 7061} - CNRS - Luminy, Case 907}

{\ten F-13288 Marseille Cedex 9 - France }

\vspace{4 cm}

{\twelve THE CONVERSION OF GRAVITONS INTO PHOTONS IN $TE_{mo}$ MODE}

\vspace{0.3 cm}
\setcounter{footnote}{0}
\renewcommand{\thefootnote}{\arabic{footnote}}

{\bf Nguyen Suan Han\footnote{Permanent address:
                 Department of Theoretical Physics, Hanoi State University,

                 \quad P. O. Box 600, BoHo, Hanoi, 10000, Vietnam}, Hoang
Ngoc Long\footnote{Institute of Theoretical Physics, P. O. Box
429, BoHo, Hanoi10000,
Vietnam.} and Dang
Van Soa$^{1}$}

\vspace{1.5 cm}

{\bf Abstract}

\end{center}

	We consider the conversion of gravitons into photons in the $
TE_{mo} $ mode.  Cross sections in different directions are given.

\vspace{4cm}

\noindent November 1994

\noindent CPT-94/P.3080

\bigskip

\noindent anonymous ftp or gopher: cpt.univ-mrs.fr

\end{titlepage}

\section{Introduction}

	The study on the interaction between electromagnetic (EM) and
   gravitational fields is a significant part of the research on
gravitational
   radiation $ [ 1-3 ] $. Based on the present technical level in
the
laboratory, it has been proved by means of mechanics that it is as
yet
   difficult to generate gravitational wave (GW) which are strong
enough to be
   detected by current detectors. Therefore physicists have
transfered their
   interest to strong EM field and taken it as one of the possible
sources in
   the laboratory.

	In previous paper $ [ 4 ] $ we have considered the conversion
of
   gravitons into photons in the periodic external EM field, namely
in the
   $TE_{10}$ mode.   In this paper we consider the conversion of
gravitons into
   photons in the $ TE_{mo}$ mode $ [ 5 ] $ :

\section{Cross sections}
	We consider the case of an alternating EM field in a
rectangular
   cavity  with inner dimensions a, b, propagating in $ z = - c/2
$ and $z=c/2$. To simplify we use a system of units in which
$\hbar=c=1$, choose
   the nontrivial solution $ TE_{mo} $ mode \cite{jackson}:

$$
H_z = H_0\cos \left (\frac{m\pi x}{a} \right ) e^{ikx -i\omega t}\ ,
$$
\begin{equation}
H_x = - \frac{ika}{m\pi} H_0 \sin \left ( \frac{m\pi x}{a} \right )
e^{ikz-i\omega t}\ ,
\end{equation}
$$
E_y = i \frac{\omega a}{m\pi} H_0 \sin \left (\frac{ m\pi x }{ a }
\right )
e^{ ikz - i\omega t }\ ,
$$

   where $ k = w/v $.

       Let us consider the conversion of the graviton $ g $ with
momentum $ p $
   into a photon $ \gamma $ with momentum $ q $ in the above
mentioned external
   EM field. For the above mentioned process we get the matrix
element:

$$
< q| M_{g\gamma}|p> = \frac{k}{4 (2\pi)^2 \sqrt{q_0p_0}}
\epsilon^i (\vec q, \sigma)
\left[ B \epsilon^{j 2} (\vec p, \tau) \eta_{ij} q_0 +
(\eta_{i2}q_j -
\eta_{ij} q_y)(D\epsilon^{j3}(\vec p, \tau) - C \epsilon^{j 1}
(\vec p, \tau))\right.
$$
\begin{equation}
\left.  +(C\eta_{i1} - D\eta_{i3}) \epsilon^{j2} (\vec p, \tau) q_j
+
\epsilon^{j2}(\vec p, \tau)\eta_{ij}(Dq_z - C q_x)\right]\ ,
\end{equation}

where $ k = \sqrt{16\pi G}, q_o = p_o +\omega $  and
$$
B = \frac{ (-1)^{m-1}/2 8a\omega H_0 (p_x - q_x )}{ m\pi [ ( p_x-
q_x )^{2}
- \frac{m^{2} \pi^{2}}{a^{2}}( p_y - q_y) (p_z -q_z
+k)}\cos\frac{a}{2}
( p_x - q_x )\sin \frac{b}{2}( p_y -q_y )\sin\frac {c}{2}( p_z -q_z
+k ),
$$
$$
C = - \frac{ (-1)^\frac{m-1}{2} 8 m \pi H_0} { a [ ( p_x - q_x )^2
- \frac{m^2\pi^2}{ a^2}]( p_y - q_y ) (p_z - q_z +k )}
\cos \frac{a}{2}( p_x - q_x )  \\ \nonumber
\sin \frac{b}{2}(p_y -q_y) \sin \frac{c}{2}( p_z - q_z +k ),
$$
$$
D = - \frac{ (-1)^{(m-1 )/2} 8kaH_0 ( p_x - q_x )}
       {m\pi [ ( p_x - q_x )^2 - \frac{m^2\pi^2}{a^2}]
       ( p_y - q_y )( p_z - q_z +k )}
\cos\frac{a}{2}\sin\frac{b}{2}(p_y - q_y )
       \sin\frac{c}{2}( p_z -q_z + k )
$$
for $ m = 2k-1, k = 1,2,3,.. .  $  and

$$
B = \frac{- (-1)^{m/2} 8\omega H_0}{[ ( p_x - q_x )^2 -
\frac{m^2\pi^2}{a^2}](p_y -q_y)(p_z - q_z + k ) }
\sin\frac{a}{2}( p_x - q_x ) sin \frac{b}{2}( p_y - q_y)
\sin\frac{c}{2}( p_z - q_z +k ),
$$
$$
C = \frac{ (-1)^{m/2} 8 H_0 (p_x - q_x )}
{ [( p_x-q_x )^2 - \frac{m^2 \pi^2}{a^2}] (p_y -q_y)(p_z - q_z + k)
}
\sin \frac{a}{2}( p_x - q_x ) \sin \frac{b}{2}( p_y - q_y )
\sin \frac{c}{2}( p_z -q_z + k ),
$$
$$
D = \frac{ ( - 1 )^{m/2} 8 k H_0}{ [ (p_x - q_x )^2 -
\frac{ m^2 \pi^2}{ a^2} ]( p_y - q_y )( p_z - q_z + k )}
\sin \frac{a}{2}(p_x - q_x)\sin \frac{b}{2}( p_y - q_y )
\sin \frac{c}{2} ( p_z - q_z + k ).
$$
   for $ m = 2k $.

       In the following, we assume that $ v = 1 $  and $ \omega<< p
$.  We also
   retain the terms to the first order in $ \omega $ in the result.

       When the momentum of the graviton is parallel to the $ z $
axis
   (the direction of the external field propagation ), the
differential
   cross section, in the context of the previous statement, vanishes
  $ (\sim O( \omega^2 ) ) $ when $ \theta \approx 0 $ and when  $
\theta = \pi/2 $.

      Now we consider the case in which the momentum of graviton is
   parallel to the $ x $ axis, i. e. $  p^{\mu} = ( p,p,0,0) $.  We
have

$$
{\frac{d\sigma ( g \rightarrow \gamma)}{d \Omega'}} =
\frac{ k^2 q}{64(2 \pi)^2 p }\left\{ q^2 C^2 ( 1 + \cos^2\theta -
2\sin^2
\theta\cos^2 \varphi' ) -\right.
$$$$
 2C \left[ q\cos \theta ( 1 + \cos^2 \theta + \sin^2 \theta) ( q_0
B + q D\sin
\theta \sin \varphi')\right.
$$
  \begin{equation} - q^2 \sin( 2 \theta) \sin
\left. \left.  \varphi\right]\right\}\ .
 \end{equation}

  From ( 3 ) it follows that $ ( m = 2k-1) $

\begin{equation}
\frac{d\sigma (g \rightarrow  \gamma)}{d\Omega'} =
\frac{k^2 {H_0}^2 V^2 p^2}{2m^2(2\pi)^4}
\left( 1 + 3\frac{\omega}{p} \right )
\end{equation}

for $ \theta \approx 0 $, and

\begin{equation}
\frac{d\sigma (g \rightarrow  \gamma) }{ d\Omega'} = \\
\frac{k^2 H_{0}^{2} b^2 m^2}{ 16a^2 ( p^2 - \frac{m^2 \pi^2 }{ a^2
})^2 }
\left( 1 +3\frac{\omega}{p} \right) \cos^2 \left(\frac{ap}{2}
\right)
2 \sin^2  [\frac{ c(p + 2k) } { 2 } ]
\end{equation}

for $ \theta = \frac{\pi}{2},\theta=\frac{\pi}{2} $ and for $ m = 2k $

\begin{equation}
\frac{d\sigma ( g \rightarrow  \gamma)}{d\Omega'} =  0
\end{equation}

   for the $ \theta  \approx 0 $, and

\begin{equation}
\frac{d\sigma (g \rightarrow  \gamma) }{ d\Omega'} =  \\
\frac{k^2 H^{2}_{0} b^2 p^2}{4(2\pi)^4(p^2-\frac{m^2\pi^2}{a^2})}
\left ( 1+ 3\frac{\omega}{p} \right)
\sin^2\left (\frac{ap}{2} \right)
2\sin^2  \left [\frac{c(p + 2k)}{2}\right]
\end{equation}
   for $  \theta = \pi/2 , \varphi = \pi/2 $.

       Next, we consider the case in which the momentum of the
graviton
   is parallel to the $ y $ axis, i.  e.  $ p^ {\mu} =(p,0,p,0) $. We
then have

\begin{equation}
\frac{d \sigma ( g \rightarrow  \gamma) }{ d \Omega''}=
\frac{k^2 q^3}{64(2\pi)^2 p} ( C^2 + D^2 ) ( 3 \cos^2\theta - 1)\ .
\end{equation}

   From (8) it follows that the differential cross section vanishes
in the
   region $ 120^0 \gg \theta\gg 60^0 $ and gets the largest values.

\begin{equation}
\frac{d\omega( g \rightarrow  \gamma)}{d\Omega''} =
\frac{k^2 {H_c}^2 V^2 p^2}{2(2\pi)^4  m^2} \left( 1 + 3
\frac{\omega}{p} \right)
\end{equation}

   for $ \theta \approx 0$ and $m=2k-1 $.

\section{Acknowledgements}
     We are grateful to Profs. B.M. Barbashov, V.N. Pervushin, A.M.
Khvedelidze, Dao Vong Duc, Tran Huu Phat for valuable discussions.
One of us (NSH) is indebted to Prof. P. Chiappetta for financial
support during his stay at the Centre de Physique Th\'eorique in
Marseille and warm hospitality.

\end{document}